\documentclass[twocolumn,prl,aps,showpacs,superscriptaddress]{revtex4}

\usepackage{graphicx}
\usepackage{color}

\begin{document}
\newcommand{\zrzn}{ZrZn$_2$}
\newcommand{\ZrZn}{ZrZn$_2$}
\newcommand{\uge}{UGe$_2$}
\newcommand{\MnSi}{MnSi\,-\,110 }
\newcommand{\Mnsi}{MnSi\,-\,111 }
\newcommand{\Tc}{$T_{c}$ }
\newcommand{\rhoxx}{$\rho_{xx}$ }
\newcommand{\rhoxy}{$\rho_{xy}$ }

\renewcommand{\floatpagefraction}{0.5}

\title{Topological Hall effect in the A-phase of MnSi}

\author{A. Neubauer}

\affiliation{Physik Department E21, Technische Universit\"at M\"unchen, James-Franck-Strasse, D-85748 Garching,
Germany}

\author{C. Pfleiderer}

\affiliation{Physik Department E21, Technische Universit\"at M\"unchen, James-Franck-Strasse, D-85748 Garching,
Germany}

\author{B. Binz}

\affiliation{Institute for Theoretical Physics, Universit\"at zu K\"oln, 50937 Cologne, Germany}

\author{A. Rosch}

\affiliation{Institute for Theoretical Physics, Universit\"at zu K\"oln, 50937 Cologne, Germany}

\author{R. Ritz}

\affiliation{Physik Department E21, Technische Universit\"at M\"unchen, James-Franck-Strasse, D-85748 Garching,
Germany}

\author{P. Niklowitz}

\affiliation{Physik Department E21, Technische Universit\"at M\"unchen, James-Franck-Strasse, D-85748 Garching,
Germany}

\author{P. B\"oni}

\affiliation{Physik Department E21, Technische Universit\"at M\"unchen, James-Franck-Strasse, D-85748 Garching,
Germany}

\date{\today}

\begin{abstract}
Recent small angle neutron scattering suggests, that the spin structure in the A-phase of MnSi is a so-called triple-$Q$ state, i.e., a superposition of three helices under 120 degrees. Model calculations suggest that this structure in fact is a lattice of so-called skyrmions, i.e., a lattice of topologically stable knots in the spin structure. We report a distinct additional contribution to the Hall effect in the temperature and magnetic field  range of the proposed skyrmion lattice, where such a contribution is neither seen nor expected for a normal helical state. Our Hall effect measurements constitute a direct observation of a topologically quantized Berry phase that identifies the spin structure seen in neutron scattering as the proposed skyrmion lattice.
\end{abstract}


\vskip2pc

\maketitle

Many years ago Skyrme showed that topologically stable objects of a nonlinear field theory for pions can be interpreted as protons or neutrons \cite{skyr61,acto79}. This seminal paper inspired the search for topological stable particle-like objects in a broad range of fields ranging from high-energy to many areas of condensed matter physics.  For instance, twenty years ago it has been predicted that skyrmions exist in anisotropic spin systems with chiral spin-orbit interactions, where they are expected to form crystalline structures \cite{bogd89,bogd94}. Lattices of skyrmions have also been suggested to occur in dense nuclear matter \cite{kleb85} or in quantum Hall systems near Landau level filling factor $\nu=1$ \cite{brey95}. However, thus far the experimental evidence is only indirect \cite{gerv05,hen08}.

Recently we reported microscopic evidence of a skyrmion lattice in the A-phase of the transition metal compound MnSi using small angle neutron scattering (SANS)  \cite{mueh08}.  The SANS data shows magnetic Bragg peaks with a hexagonal symmetry consistent with the superposition of three helices under an angle of 120 degrees -- a so-called triple-$Q$ structure. The three helices are thereby confined to a plane strictly perpendicular to the applied magnetic field. A detailed theoretical analysis \cite{mueh08} of an appropriate Ginzburg-Landau model suggested that a lattice of anti-skyrmion lines forms in the A-phase of MnSi, similar to the vortex lattice in superconductors.

However, whether the spin structure in the A-phase indeed represents a skyrmion lattice depends crucially on the phase relationship of the helices that are superimposed. This phase information could not be extracted from the SANS data. In contrast to neutron scattering the phase relationship of the helices, and thus existence of topologically nontrivial spin structures, may be established directly by means of the so-called topological Hall effect (THE) \cite{binz08}. The perhaps most convincing example of a topological Hall effect has been reported for 3D pyrochlore lattices \cite{tagu01,machida07}. However, in these systems the non-coplanar spin structure is due to frustration on short length scales, i.e., the spin structure is not a continuous field for which topological properties may be defined in a strict sense. The topological Hall effect has also been considered, e.g., in La$_{1-x}$Co$_x$MnO$_3$ \cite{ye99}, CrO$_2$ \cite{yana02}, and Gd \cite{bail05}, but there is essentially no independent microscopic information on the relevant spin structures.  

The origin of the topological Hall effect is a Berry phase collected by the conduction electrons when following adiabatically the spin polarization of topologically stable knots in the spin structure \cite{binz08,ye99,tagu01,onod04,brun04,tata07}. Thus the Berry phase reflects the chirality and winding number of the knots. The topological Hall effect arises besides the normal Hall effect, which is proportional to the applied magnetic field, and the anomalous Hall effect (AHE) that scales with ferromagnetic components of the magnetization \cite{naga06,karp54,smit58}. The AHE may be viewed in terms of Berry curvature in momentum space  \cite{onod02,jungwirth02} as opposed to real space for the topological Hall effect.

In our study of the Hall effect in MnSi we find a distinct anomalous contribution in the A-phase. The sign of this contribution is opposite to the normal Hall effect and the prefactor is quantitatively consistent with the skyrmion density derived from neutron scattering and theory. The observation of this Hall effect provides clear experimental evidence that the magnetic structure observed in neutron scattering has indeed the topological properties (chirality and winding number) of the proposed skyrmion lattice.

As a function of temperature $T$ the itinerant-electron magnet MnSi develops long-range single-$Q$ helimagnetic order below $T_c=29.5\,{\rm K}$. The helimagnetic state may be understood as the result of a hierarchy of energy scales \cite{naka80,bak80}, where ferromagnetic exchange on the strongest scale and the isotropic Dzyaloshinsky-Moriya spin-orbit interactions due to the lack of inversion symmetry of the cubic B20 structure give rise to a long-wavelength helimagnetic modulation, where $\lambda_h\approx190\,{\rm \AA}$. The propagation vector $\vec{Q}$ of the helix is pinned to the cubic space-diagonal, $\vec{Q}\parallel[111]$, by higher order spin-orbit coupling terms, which represent the weakest scale.

Under magnetic field the helical wave vector $\vec{Q}$ is unpinned from the $[111]$ direction and aligns parallel to the applied magnetic field for $B>B_{c1}\approx0.1\,{\rm T}$. The magnetic state above $B_{c1}$ is also referred to as a conical state, because it consists of a superposition of a helical modulation $\vec{M}_Q$ with a uniform magnetization $\vec{M}_0$, where $\vec{Q}\parallel\vec{M}_0$. The helical modulation is suppressed altogether for a magnetic field exceeding $B_{c2}(T\to0)\approx0.6\,{\rm  T}$. In the vicinity of $T_c$ a small phase pocket has been observed referred to as the A-phase (see also Fig.\,\ref{figure4}) \cite{ishi84}. The specific heat, susceptibility and neutron scattering establish, that the A-phase is a distinct phase with a first order phase transition separating it from the conical phase. It had further been established that $\vec{Q}$ in the A-phase aligns perpendicular to the applied magnetic field \cite{lebe93,grig06b}, however, neither the full spin structure had been resolved, nor was there a plausible explanation for $\vec{Q}\perp\vec{B}$ prior to our study \cite{mueh08}.

\begin{figure}
\includegraphics[width=0.8\linewidth,clip=]{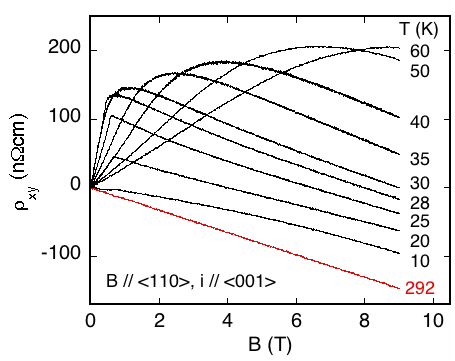}
\caption{Hall resistivity for single crystal MnSi, where magnetic field $B$ was applied parallel to $[110]$ and current was applied along $[001]$. Data for magnetic field $B\parallel[111]$ and current $I\parallel[1\bar{1}0]$ (not shown) are the same, as expected for a cubic material.} \label{figure1}
\end{figure}

The Hall effect and the magnetoesistance in MnSi have been studied before for temperatures below $T_c$ and magnetic field up to 5\,T \cite{lee07}. These measurements were analyzed in terms of the sum of normal and anomalous Hall currents, $\sigma_{xy}=\sigma_{xy}^N+\sigma_{xy}^A$, respectively. This Ansatz contrasts the conventional Karplus-Luttinger Ansatz of a sum of normal and anomalous Hall resistivities, $\rho_{xy}=R_0B+\mu_0 R_sM$. It was in particular noticed that below $T_c$,  $\sigma_{xy}^A=S_H\,M$, where $S_H$ is independent of $T$ and $B$ while $\sigma_{xy}^N\approx -R_0 B/\rho_{xx}^2$ changes by a factor of 100 between 5\,K and $T_c$, reflecting the strong $T$-dependence of the resistivity $\rho_{xx}$.

For our study single-crystal samples were cut from an ingot that had been studied before by various bulk properties, SANS \cite{mueh08} and Larmor diffraction \cite{pfle07}. The samples were oriented with x-ray Laue diffraction and polished to size. Sample 1 was oriented for measurements with $B$ parallel $[110]$ and electric current $I$ parallel $[001]$ and sample 2 for $\vec{B}\parallel[111]$ and $I\parallel[1\bar{1}0]$. The sample dimensions as determined with a light microscope were $1\times 1.5\times 0.13\,{\rm mm^{3}}$ and $1.6\times 3.1\times 0.15\,{\rm mm^{3}}$ for sample 1 and 2, respectively. Quite generally the geometry factor in studies of this kind can be determined only quite inaccurately. Because MnSi has a cubic structure and $T_c$ is small as compared with room temperature we determined the geometry factors from the longitudinal and Hall resistivities at ambient conditions, $\rho_{xx}{\rm  (300K,0T)}=180\,{\rm\mu\Omega cm}$ and $\rho_{xy}{\rm (300K,  8T)}=-126\,{\rm n\Omega cm}$, respectively \cite{pfle97,neub06,lee07} (note the difference of units for $\rho_{xy}$). Data reported in this paper were corrected for demagnetizing effects, where the demagnetizing factors were determined consistently from measurements of the DC magnetization for various sample dimensions and theoretical estimates.

\begin{figure}
\includegraphics[width=0.8\linewidth,clip=]{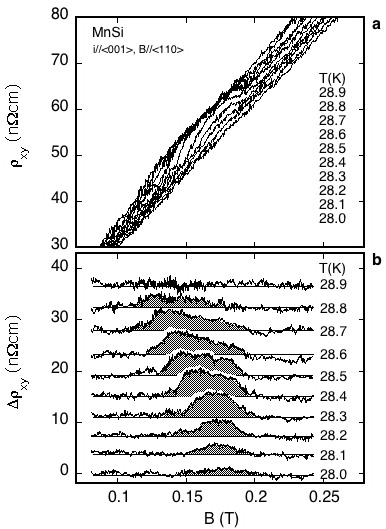}
\caption{(a) Hall resistivity $\rho_{xy}$ near $T_c$ in the temperature and field range of the A-phase. (b)
Additional Hall contribution $\Delta\rho_{xy}$ in the A-phase. Data are shifted vertically for better
visibility.} \label{figure2}
\end{figure}

The resistivity and the Hall effect were measured simultaneously in a standard six terminal configuration. Data were recorded down to 2.5\,K at magnetic fields up to 9\,T.  Symmetric and antisymmetric signal contributions in $\pm B$ were determined, where data shown here for $\rho_{xy}$ represent the antisymmetric part of the signal at the Hall contacts. We note that our Hall data are perfectly consistent with previous studies \cite{lee07}. However, we have achieved a much better resolution, making possible the observation of the additional anomalous contributions in the A-phase (for further details see Ref.\,\cite{neub06}).

Shown in Fig.\,\ref{figure1} is the Hall resistivity $\rho_{xy}$ of MnSi for $\vec{B}\parallel[110]$ and $I\parallel[001]$. At room temperature the behavior is dominated by the normal Hall effect, where we observe essentially no $T$ dependence. In the conventional interpretation the slope of the Hall resistivity corresponds to a nominal charge carrier concentration $n=(R_0 e)^{-1} =3.78\cdot10^{22}\,{\rm cm^{-3}}$ \cite{comment-lee07}. The overall behavior of $\rho_{xy}$ at low $T$ is fairly complex, but perfectly consistent with Ref.\,\cite{lee07}. 

\begin{figure}
\includegraphics[width=0.8\linewidth,clip=]{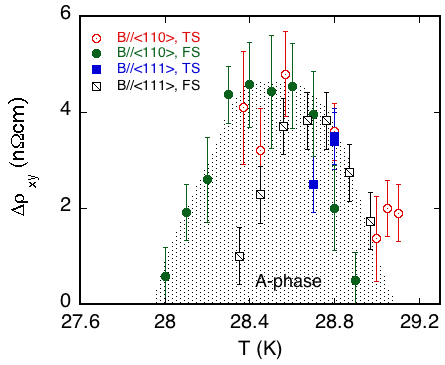}
\caption{Estimated contribution to the Hall effect in the A-phase. FS denotes data obtained in field sweeps; TS
denotes data obtained in temperature sweeps.} \label{figure3}
\end{figure}

\begin{figure}
\includegraphics[width=0.8\linewidth,clip=]{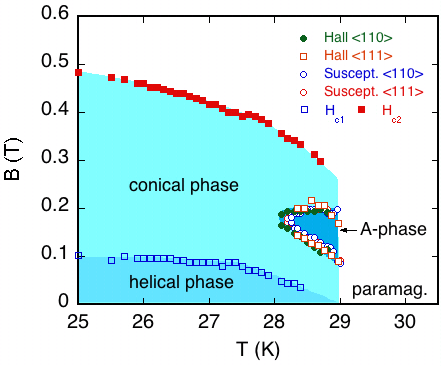}
\caption{Magnetic phase diagram of MnSi. Comparison of the A-phase as measured in the AC susceptibility versus
the Hall resistivity. Data points of the AC susceptibility are taken from Ref.\,\cite{lama06,thes97}.}
\label{figure4}
\end{figure}

Shown in Fig.\,\ref{figure2}(a) is $\rho_{xy}$ as measured in the regime of the A-phase, where a small additional contribution appears. We have approximated the signal linearly from below to above the A-phase and subtracted this part of the total signal. The resulting contribution $\Delta\rho_{xy}$ is shown in Fig.\,\ref{figure2}(b), where the curves have been shifted vertically for better visibility \cite{slope}. In Fig.\,\ref{figure3} we show a rough estimate of the magnitude of the contribution, where we plot the peak value. The error bars represent a conservative estimate of systematic errors. Within experimental uncertainties we find $\Delta\rho_{xy}\approx4.5\pm1\,{\rm n \Omega cm}$.

As a final test we find remarkable agreement between the field and $T$ range in which we observe $\Delta\rho_{xy}$ with the regime of the A-phase inferred from the AC susceptibility reported in \cite{lama06} (Fig.\,\ref{figure4}). This clearly confirms that the additional Hall signal is correctly attributed to the A-phase. The key features of $\Delta\rho_{xy}$ observed in the A-phase may be summarized as follows: (i) the sign of the signal is opposite to the normal Hall effect; (ii) the magnitude of the signal is roughly  $\Delta\rho_{xy}\approx4.5\,{\rm n\Omega cm}$, (iii) the signal is roughly the same for $\vec{B}\parallel[110]$ and $\vec{B}\parallel[111]$ and thus essentially independent of direction.

We note, that the magnetization in the A-phase does not show an additional ferromagnetic contribution that would explain the additional anomalous contribution. Instead, for increasing $B$ the magnetization slightly increases both when entering and when leaving the A-phase at $B_{a1}$ and $B_{a2}$, respectively \cite{greg92,thes97,lama06}. Correspondingly the slope of $M(B)$ is reduced in the A-phase. Thus, $\Delta\rho_{xy}$ must be related to the modulated spin structure observed in neutron scattering \cite{mueh08}. This already establishes the existence of topologically stable knots in the spin structure.

Further, when the motion of conduction electrons follows a topologically non-trivial spin structure, the charge carriers collect a Berry phase. This Berry phase may be viewed as an Aharonov-Bohm phase arising from a fictitious effective  field $\vec{B}_{\rm eff}=\Phi_0\,\vec{\Phi}$ with opposite sign for majority and minority spins, where $\Phi_0=h/e$ is the flux quantum for a single electron \cite{ye99,brun04,tata07,binz08}. Here $\vec{\Phi}$
is given by the skyrmion density
\begin{equation}
\Phi^\mu=\frac{1}{8\pi} \epsilon_{\mu\nu\lambda}
\hat{n}\cdot(\partial_{\nu}\hat{n}\times\partial_{\lambda}\hat{n}) \label{skyrm}
\end{equation}
where $\epsilon_{\mu\nu\lambda}$ is the antisymmetric unit tensor and $\hat{n}=\vec{M}/\vert M \vert$ \cite{comment-machida07}. The integrated skyrmion density per unit cell is a measure for a winding number and is therefore quantized to an integer.  

As for the normal Hall effect, the precise value of the topological Hall contribution due to $\vec{B}_{\rm eff}$ depends in a multi-band system like MnSi on details of the band structure and the relative size of scattering rates. Because these factors also enter in $R_0$ in a similar way, using the measured value of $R_0$ in Eq.~(\ref{hall}) allows for a semi-quantitative prediction. In the adiabatic limit, where the spin-polarization of charge carriers with infinite lifetime smoothly follows the texture $\vec{M}$, the topological Hall signal may be expressed as \cite{brun04,tata07}
\begin{equation}
\Delta\rho_{xy}\approx P\,R_0\,B^z_{\rm eff} \label{hall}
\end{equation}
where $\hat{z}$ is the direction of the applied field, $R_0$ is the normal Hall constant given above and $P$ measures the local spin-polarization of the conduction electrons. The factor $P$ arises as majority- and minority-spin carriers collect Berry phases of opposite sign. Therefore the signal vanishes for vanishing polarization, $P\to0$ and is maximal for a fully polarized system, $P=1$. For a single-$Q$ structure $\vec{\Phi}=0$ so that $\vec{B}_{\rm eff}=0$ and there is no topological Hall effect. 
  
We may now compare the experimentally observed Hall voltage $\Delta\rho_{xy}\approx4.5\,{\rm n\Omega m}$ with the predicted topological Hall signal. For the proposed lattice of anti-skyrmion lines in the A-phase of MnSi $\int\!dx\,dy\,\Phi^{z}=-1$ for each 2-dimensional magnetic unit cell \cite{mueh08}. This implies that the effective field is quantized and oriented opposite to the applied field. For MnSi it follows that $B_{\rm eff}\approx 2.5\,{\rm  T}$. The polarization $P=\mu_{\rm spo}/\mu_{\rm sat}$ represents the ratio of the ordered magnetic moment in the A-phase $\mu_{\rm spo}\approx0.2\pm0.05\,{\rm \mu_B}$ to the saturated magnetic moment $\mu_{sat}\approx2.2\pm0.2\,{\rm \mu_B}$ where the saturated moment may be taken, e.g., from the Curie Weiss moment in the paramagnetic state or the free Mn moment \cite{bloc75}. Hence the polarization is given by $P\approx0.1\pm0.02$. Taken together the theoretically predicted value of $\Delta\rho_{xy}\approx 4\,{\rm n\Omega cm}$ is in remarkable agreement with experiment.

While writing this manuscript a similar Hall effect has been reported for MnSi in the pressure range 6 to 12\,kbar \cite{lee08}. This signal is over an order of magnitude larger than the signal we report here and extends over a much larger field range (0.1 and 0.5\,T). Detailed susceptibility and magnetization measurements under pressure reported, e.g.,  in Ref.\,\cite{thes97} indicate, that the magnetic field range of the A-phase and the conical phase are unchanged up to roughly 11 kbar, where the A-phase seems to vanish. Hence, due to the lack of neutron scattering data under magnetic field and pressure in the range 6 to 12\,kbar the precise spin structure represents an exciting question for future research.

In conclusion, when taking together the sign and quantitative size of $\Delta\rho_{xy}$ with the neutron scattering data reported in Ref.\,\cite{mueh08}, our study identifies the A-phase of MnSi as the proposed lattice of skyrmions. In fact, our Hall effect data consitute a direct observation of a topologically quantized Berry phase, thereby unambiguously identifying the proposed spin structure inferred from neutron scattering.

We wish to thank A. Bauer, R. Duine, M. Garst, F. Jonietz, S. Legl, S. M\"uhlbauer, W. M\"unzer, B. Russ, J. Schmalian, M. Vojta and A. Vishwanath and gratefully acknowledge financial support by SFB608 and the Alexander-von-Humboldt foundation.


\begin{thebibliography}{41}
\expandafter\ifx\csname natexlab\endcsname\relax\def\natexlab#1{#1}\fi
\expandafter\ifx\csname bibnamefont\endcsname\relax
  \def\bibnamefont#1{#1}\fi
\expandafter\ifx\csname bibfnamefont\endcsname\relax
  \def\bibfnamefont#1{#1}\fi
\expandafter\ifx\csname citenamefont\endcsname\relax
  \def\citenamefont#1{#1}\fi
\expandafter\ifx\csname url\endcsname\relax
  \def\url#1{\texttt{#1}}\fi
\expandafter\ifx\csname urlprefix\endcsname\relax\def\urlprefix{URL }\fi
\providecommand{\bibinfo}[2]{#2}
\providecommand{\eprint}[2][]{\url{#2}}

\bibitem[{\citenamefont{Skyrme}(1961)}]{skyr61}
\bibinfo{author}{\bibfnamefont{T.~H.} \bibnamefont{Skyrme}},
  \bibinfo{journal}{Proc. Roy. Soc. Lond. A}
  \textbf{\bibinfo{volume}{260}}, \bibinfo{pages}{127} (\bibinfo{year}{1961}).

\bibitem[{\citenamefont{Actor}(1979)}]{acto79}
\bibinfo{author}{\bibfnamefont{A.}~\bibnamefont{Actor}},
  \bibinfo{journal}{Rev.~Mod.~Phys.} \textbf{\bibinfo{volume}{51}},
  \bibinfo{pages}{461} (\bibinfo{year}{1979}).

\bibitem[{\citenamefont{Bogdanov and Yablonskii}(1989)}]{bogd89}
\bibinfo{author}{\bibfnamefont{A.~N.} \bibnamefont{Bogdanov}} \bibnamefont{and}
  \bibinfo{author}{\bibfnamefont{D.~A.} \bibnamefont{Yablonskii}},
  \bibinfo{journal}{Sov.\ Phys.\ JETP} \textbf{\bibinfo{volume}{68}},
  \bibinfo{pages}{101} (\bibinfo{year}{1989}).

\bibitem[{\citenamefont{Bogdanov and Hubert}(1994)}]{bogd94}
\bibinfo{author}{\bibfnamefont{A.}~\bibnamefont{Bogdanov}} \bibnamefont{and}
  \bibinfo{author}{\bibfnamefont{A.}~\bibnamefont{Hubert}},
  \bibinfo{journal}{J. Magn. Magn. Mater.} \textbf{\bibinfo{volume}{138}},
  \bibinfo{pages}{255} (\bibinfo{year}{1994}).

\bibitem[{\citenamefont{Klebanov}(1985)}]{kleb85}
\bibinfo{author}{\bibfnamefont{I.}~\bibnamefont{Klebanov}},
  \bibinfo{journal}{Nucl. Phys. B} \textbf{\bibinfo{volume}{262}},
  \bibinfo{pages}{133} (\bibinfo{year}{1985}).

\bibitem[{\citenamefont{Brey et~al.}(1995)\citenamefont{Brey, Fertig, Cote, and
  MacDonald}}]{brey95}
\bibinfo{author}{\bibfnamefont{L.}~\bibnamefont{Brey}},
  \bibinfo{author}{\bibfnamefont{et} \bibnamefont{al.}},
  \bibinfo{journal}{Phys. Rev. Lett.} \textbf{\bibinfo{volume}{75}},
  \bibinfo{pages}{2562} (\bibinfo{year}{1995}).

\bibitem[{\citenamefont{Gervais et~al.}(2005)\citenamefont{Gervais, Stormer,
  Tsui, Kuhns, Moulton, Reyes, Pfeiffer, Baldwin, and West}}]{gerv05}
\bibinfo{author}{\bibfnamefont{G.}~\bibnamefont{Gervais}},
  \bibinfo{author}{\bibfnamefont{et} \bibnamefont{al.}},
  \bibinfo{journal}{Phys. Rev. Lett.} \textbf{\bibinfo{volume}{94}},
  \bibinfo{pages}{196803} (\bibinfo{year}{2005}).

\bibitem[{\citenamefont{Hen and Karliner}(2008)}]{hen08}
\bibinfo{author}{\bibfnamefont{I.}~\bibnamefont{Hen}} \bibnamefont{and}
  \bibinfo{author}{\bibfnamefont{M.}~\bibnamefont{Karliner}},
  \bibinfo{journal}{Phys. Rev. D} \textbf{\bibinfo{volume}{77}},
  \bibinfo{eid}{054009} (\bibinfo{year}{2008}).

\bibitem[{\citenamefont{M\"uhlbauer et~al.}(2009)\citenamefont{M\"uhlbauer,
  Binz, Jonietz, Pfleiderer, Rosch, Neubauer, Georgii, and B\"oni}}]{mueh08}
\bibinfo{author}{\bibfnamefont{S.}~\bibnamefont{M\"uhlbauer}},
  \bibinfo{author}{\bibfnamefont{et}~\bibnamefont{al.}},
  \bibinfo{journal}{Science} \textbf{\bibinfo{volume}{323}},
  \bibinfo{pages}{915} (\bibinfo{year}{2009}).

\bibitem[{\citenamefont{Binz and Vishwanath}(2008)}]{binz08}
\bibinfo{author}{\bibfnamefont{B.}~\bibnamefont{Binz}} \bibnamefont{and}
  \bibinfo{author}{\bibfnamefont{A.}~\bibnamefont{Vishwanath}},
  \bibinfo{journal}{Physica B} \textbf{\bibinfo{volume}{403}},
  \bibinfo{pages}{1336} (\bibinfo{year}{2008}).

\bibitem[{\citenamefont{Taguchi et~al.}(2001)\citenamefont{Taguchi, Oohara,
  Yoshizawa, Nagaosa, and Tokura}}]{tagu01}
\bibinfo{author}{\bibfnamefont{Y.}~\bibnamefont{Taguchi}},
  \bibinfo{author}{\bibfnamefont{et}~\bibnamefont{al.}},
  \bibinfo{journal}{Science} \textbf{\bibinfo{volume}{291}},
  \bibinfo{pages}{2573} (\bibinfo{year}{2001}).

\bibitem[{\citenamefont{Machida et~al.}(2007)
\citenamefont{Machida, Nakatsuji, Maeno, Tayama, Sakakibara, and Onoda}}]{machida07}
\bibinfo{author}{\bibfnamefont{Y.}~\bibnamefont{Machida}},
  \bibinfo{author}{\bibfnamefont{et}~\bibnamefont{al.}},
  \bibinfo{journal}{Phys.\ Rev.\ Lett.} \textbf{\bibinfo{volume}{98}},
  \bibinfo{pages}{057203} (\bibinfo{year}{2007}).

\bibitem[{\citenamefont{Ye et~al.}(1999)\citenamefont{Ye, Kim, Millis,
  Shraiman, Majumdar, and Te\ifmmode \check{s}\else
  \v{s}\fi{}anovi\ifmmode~\acute{c}\else \'{c}\fi{}}}]{ye99}
\bibinfo{author}{\bibfnamefont{J.}~\bibnamefont{Ye}},
  \bibinfo{author}{\bibfnamefont{et} \bibnamefont{al.}},
 \bibinfo{journal}{Phys. Rev. Lett.} \textbf{\bibinfo{volume}{83}}, 
 \bibinfo{pages}{3737} (\bibinfo{year}{1999}).

\bibitem[{\citenamefont{Yanagihara and Salamon}(2002)}]{yana02}
\bibinfo{author}{\bibfnamefont{H.}~\bibnamefont{Yanagihara}} \bibnamefont{and}
  \bibinfo{author}{\bibfnamefont{M.~B.} \bibnamefont{Salamon}},
  \bibinfo{journal}{Phys. Rev. Lett.} \textbf{\bibinfo{volume}{89}},
  \bibinfo{pages}{187201} (\bibinfo{year}{2002}).

\bibitem[{\citenamefont{Baily and Salamon}(2005)}]{bail05}
\bibinfo{author}{\bibfnamefont{S.~A.} \bibnamefont{Baily}} \bibnamefont{and}
  \bibinfo{author}{\bibfnamefont{M.~B.} \bibnamefont{Salamon}},
  \bibinfo{journal}{Phys. Rev. B} \textbf{\bibinfo{volume}{71}},
  \bibinfo{eid}{104407} (\bibinfo{year}{2005}).

\bibitem[{\citenamefont{Onoda et~al.}(2004)\citenamefont{Onoda, Tatara, and
  Nagaosa}}]{onod04}
\bibinfo{author}{\bibfnamefont{M.}~\bibnamefont{Onoda}},
  \bibinfo{author}{\bibfnamefont{et}~\bibnamefont{al.}},
  \bibinfo{journal}{J. Phys. Soc. Jpn.} \textbf{\bibinfo{volume}{73}},
  \bibinfo{pages}{2624} (\bibinfo{year}{2004}).

\bibitem[{\citenamefont{Bruno et~al.}(2004)\citenamefont{Bruno, Dugaev, and
  Taillefumier}}]{brun04}
\bibinfo{author}{\bibfnamefont{P.}~\bibnamefont{Bruno}},
  \bibinfo{author}{\bibfnamefont{et} \bibnamefont{al.}},
  \bibinfo{journal}{Phys. Rev. Lett.} \textbf{\bibinfo{volume}{93}},
  \bibinfo{pages}{096806} (\bibinfo{year}{2004}).

\bibitem[{\citenamefont{Tatara et~al.}(2007)\citenamefont{Tatara, Kohno,
  Shibata, Lemaho, and Lee}}]{tata07}
\bibinfo{author}{\bibfnamefont{G.}~\bibnamefont{Tatara}},
  \bibinfo{author}{\bibfnamefont{et}~\bibnamefont{al.}},
  \bibinfo{journal}{J. Phys. Soc. Jpn.} \textbf{\bibinfo{volume}{76}},
  \bibinfo{pages}{054707} (\bibinfo{year}{2007}).

\bibitem[{\citenamefont{Nagaosa}(2006)}]{naga06}
\bibinfo{author}{\bibfnamefont{N.}~\bibnamefont{Nagaosa}}, \bibinfo{journal}{J.
  Phys. Soc. Jpn.} \textbf{\bibinfo{volume}{75}}, \bibinfo{pages}{042001}
  (\bibinfo{year}{2006}).

\bibitem[{\citenamefont{Karplus and Luttinger}(1954)}]{karp54}
\bibinfo{author}{\bibfnamefont{R.}~\bibnamefont{Karplus}} \bibnamefont{and}
  \bibinfo{author}{\bibfnamefont{J.~M.} \bibnamefont{Luttinger}},
  \bibinfo{journal}{Phys. Rev.} \textbf{\bibinfo{volume}{95}},
  \bibinfo{pages}{1154} (\bibinfo{year}{1954}).

\bibitem[{\citenamefont{Smit}(1958)}]{smit58}
\bibinfo{author}{\bibfnamefont{A.~W.} \bibnamefont{Smit}},
  \bibinfo{journal}{Physica} \textbf{\bibinfo{volume}{24}}, \bibinfo{pages}{39}
  (\bibinfo{year}{1958}).

\bibitem[{\citenamefont{Onoda and Nagaosa}(2002)}]{onod02}
\bibinfo{author}{\bibfnamefont{M.}~\bibnamefont{Onoda}} \bibnamefont{and}
  \bibinfo{author}{\bibfnamefont{N.}~\bibnamefont{Nagaosa}},
  \bibinfo{journal}{J. Phys. Soc. Jpn.} \textbf{\bibinfo{volume}{71}},
  \bibinfo{pages}{19} (\bibinfo{year}{2002}).

\bibitem[{\citenamefont{Jungwirth et~al.}(2002)\citenamefont{Jungwirth, Niu,
  and MacDonald}}]{jungwirth02}
\bibinfo{author}{\bibfnamefont{T.}~\bibnamefont{Jungwirth}},
  \bibinfo{author}{\bibfnamefont{et}~\bibnamefont{al.}}, 
  \bibinfo{journal}{Phys. Rev. Lett.} \textbf{\bibinfo{volume}{88}},
  \bibinfo{pages}{207208} (\bibinfo{year}{2002}).

\bibitem[{\citenamefont{Nakanishi et~al.}(1980)\citenamefont{Nakanishi, Yanase,
  Hasegawa, and Kataoka}}]{naka80}
\bibinfo{author}{\bibfnamefont{O.}~\bibnamefont{Nakanishi}},
  \bibinfo{author}{\bibfnamefont{A.}~\bibnamefont{Yanase}},
  \bibinfo{author}{\bibfnamefont{A.}~\bibnamefont{Hasegawa}}, \bibnamefont{and}
  \bibinfo{author}{\bibfnamefont{M.}~\bibnamefont{Kataoka}},
  \bibinfo{journal}{Solid State Communi.} \textbf{\bibinfo{volume}{35}},
  \bibinfo{pages}{995} (\bibinfo{year}{1980}).

\bibitem[{\citenamefont{B{\aa}k and Jensen}(1980)}]{bak80}
\bibinfo{author}{\bibfnamefont{P.}~\bibnamefont{B{\aa}k}} \bibnamefont{and}
  \bibinfo{author}{\bibfnamefont{M.~H.} \bibnamefont{Jensen}},
  \bibinfo{journal}{J. Phys. C: Solid State} \textbf{\bibinfo{volume}{13}},
  \bibinfo{pages}{L881} (\bibinfo{year}{1980}).

\bibitem[{\citenamefont{Ishikawa and Arai}(1984)}]{ishi84}
\bibinfo{author}{\bibfnamefont{Y.}~\bibnamefont{Ishikawa}},
  \bibinfo{author}{\bibfnamefont{M.}~\bibnamefont{Arai}}, \bibinfo{journal}{J.
  Phys. Soc. Jpn.} \textbf{\bibinfo{volume}{53}}, \bibinfo{pages}{2726}
  (\bibinfo{year}{1984}).

\bibitem[{\citenamefont{Lebech}(1993)}]{lebe93}
\bibinfo{author}{\bibfnamefont{B.}~\bibnamefont{Lebech}},
  \bibinfo{journal}{Recent Adv. in Magn. of Transition Metal Comp., World
  Scientific, Singapore} p. \bibinfo{pages}{167} (\bibinfo{year}{1993}).

\bibitem[{\citenamefont{Grigoriev et~al.}(2006)\citenamefont{Grigoriev,
  Maleyev, Okorokov, Chetverikov, and Eckerlebe}}]{grig06b}
\bibinfo{author}{\bibfnamefont{S.~V.} \bibnamefont{Grigoriev}},
  \bibinfo{author}{\bibfnamefont{et} \bibnamefont{al.}},
  \bibinfo{journal}{Phys. Rev. B} \textbf{\bibinfo{volume}{73}},
  \bibinfo{eid}{224440} (\bibinfo{year}{2006}).

\bibitem[{\citenamefont{Lee et~al.}(2007)\citenamefont{Lee, Onose, Tokura, and
  Ong}}]{lee07}
\bibinfo{author}{\bibfnamefont{M.}~\bibnamefont{Lee}},
  \bibinfo{author}{\bibfnamefont{et}~\bibnamefont{al.}},
  \bibinfo{journal}{Phys. Rev. B} \textbf{\bibinfo{volume}{75}},
  \bibinfo{eid}{172403} (\bibinfo{year}{2007}).

\bibitem[{\citenamefont{Pfleiderer et~al.}(2007)\citenamefont{Pfleiderer,
  B\"oni, Keller, R\"o{\ss}ler, and Rosch}}]{pfle07}
\bibinfo{author}{\bibfnamefont{C.}~\bibnamefont{Pfleiderer}},
  \bibinfo{author}{\bibfnamefont{et}~\bibnamefont{al.}},
  \bibinfo{journal}{Science} \textbf{\bibinfo{volume}{316}},
  \bibinfo{pages}{1871} (\bibinfo{year}{2007}).

\bibitem[{\citenamefont{Pfleiderer et~al.}(1997)\citenamefont{Pfleiderer,
  McMullan, Julian, and Lonzarich}}]{pfle97}
\bibinfo{author}{\bibfnamefont{C.}~\bibnamefont{Pfleiderer}},
  \bibinfo{author}{\bibfnamefont{et} \bibnamefont{al.}},
 \bibinfo{journal}{Phys. Rev. B}
  \textbf{\bibinfo{volume}{55}}, \bibinfo{pages}{8330} (\bibinfo{year}{1997}).

\bibitem[{\citenamefont{Neubauer}(2006)}]{neub06}
\bibinfo{author}{\bibfnamefont{A.}~\bibnamefont{Neubauer}}, Master's thesis,
  \bibinfo{school}{Technische Universit\"at M\"unchen} (\bibinfo{year}{2006}).

\bibitem[{com({\natexlab{a}})}]{comment-lee07}
\bibinfo{note}{In Ref.\,\cite{lee07} the normal Hall effect was determined
  below $T_c$ in a fit of the complete Hall signal, where
  $n=8.5\cdot10^{22}\,{\rm cm^{-3}}$. We find the same value when analyzing our
  data the same way. The exact value of $n$ does not change the conclusions
  presented here.}

\bibitem[{slo()}]{slope}
\bibinfo{note}{The small negative slope of $\Delta\rho_{xy}(B)$ is possibly due
  to systematic error in the subtraction of the ordinary AHE.}

\bibitem[{\citenamefont{Lamago et~al.}(2006)\citenamefont{Lamago, Georgii,
  Pfleiderer, and B\"oni}}]{lama06}
\bibinfo{author}{\bibfnamefont{D.}~\bibnamefont{Lamago}},
  \bibinfo{author}{\bibfnamefont{et}~\bibnamefont{al.}},
  \bibinfo{journal}{Physica B} \textbf{\bibinfo{volume}{385--386}},
  \bibinfo{pages}{385} (\bibinfo{year}{2006}).

\bibitem[{\citenamefont{Gregory et~al.}(1992)\citenamefont{Gregory, Lambrick,
  and Bernhoeft}}]{greg92}
\bibinfo{author}{\bibfnamefont{C.}~\bibnamefont{Gregory}},
  \bibinfo{author}{\bibfnamefont{D.}~\bibnamefont{Lambrick}}, \bibnamefont{and}
  \bibinfo{author}{\bibfnamefont{N.}~\bibnamefont{Bernhoeft}},
  \bibinfo{journal}{J. Magn. Magn. Materials}
  \textbf{\bibinfo{volume}{104-107}}, \bibinfo{pages}{689}
  (\bibinfo{year}{1992}).

\bibitem[{\citenamefont{Thessieu et~al.}(1997)\citenamefont{Thessieu,
  Pfleiderer, Stepanov, and Flouquet}}]{thes97}
\bibinfo{author}{\bibfnamefont{C.}~\bibnamefont{Thessieu}},
  \bibinfo{author}{\bibfnamefont{C.}~\bibnamefont{Pfleiderer}},
  \bibinfo{author}{\bibfnamefont{A.~N.} \bibnamefont{Stepanov}},
  \bibnamefont{and} \bibinfo{author}{\bibfnamefont{J.}~\bibnamefont{Flouquet}},
  \bibinfo{journal}{J. Phys.: Condens. Matter} \textbf{\bibinfo{volume}{9}},
  \bibinfo{pages}{6677} (\bibinfo{year}{1997}).

\bibitem[{com({\natexlab{b}})}]{comment-machida07}
\bibinfo{note}{The effective field defined in Ref.\,\cite{machida07} for a
  discrete lattice of spins limits to the definition of the topological field
  for a continuous magnetization used here (cf. Eqn.\,\ref{skyrm}).}

\bibitem[{\citenamefont{Bloch et~al.}(1975)\citenamefont{Bloch, Voiron,
  Jaccarino, and Wernick}}]{bloc75}
\bibinfo{author}{\bibfnamefont{D.}~\bibnamefont{Bloch}},
  \bibinfo{author}{\bibfnamefont{et}~\bibnamefont{al.}},
 \bibinfo{journal}{Phys. Lett. A}
  \textbf{\bibinfo{volume}{51}}, \bibinfo{pages}{259} (\bibinfo{year}{1975}).

\bibitem[{lee()}]{lee08}
\bibinfo{author}{\bibfnamefont{M.}~\bibnamefont{Lee}},
  \bibinfo{author}{\bibfnamefont{et}~\bibnamefont{al.}},
  \bibinfo{journal}{Phys.\ Rev.\ Lett.} \textbf{\bibinfo{volume}{102}},
  \bibinfo{pages}{186601} (\bibinfo{year}{2009});
	\bibinfo{note}{arXiv/08113146.}

\end{thebibliography}

\end{document}